\documentclass[pra,showpacs,twocolumn,superscriptaddress,floatfix,nofootinbib,aps,longbibliography]{revtex4-2}
\usepackage{amsmath,amssymb,bm}
\usepackage{graphicx}
\usepackage{physics}
\usepackage[T1]{fontenc}
\usepackage{comment}
\usepackage{float}
\usepackage{mathtools}
\usepackage[normalem]{ulem}  
\usepackage{hyperref}
\usepackage{soul}
\usepackage{braket}
\usepackage[utf8]{inputenc}
\hypersetup{colorlinks,%,%
 linkcolor=blue,%
 citecolor=blue,%
 urlcolor=blue}
\usepackage{color}
\usepackage{xcolor}
\usepackage{needspace}

\sethlcolor{yellow}

\usepackage[dvipsnames]{xcolor}
\usepackage{tikz}
\usetikzlibrary{shapes}
\usepackage{placeins}

\begin{document}

\title{Faraday pattern formation in dipolar superfluid and supersolid quantum gases}

\author{Sumit Semwal}
\affiliation{Department of Physics, Indian Institute of Technology, Guwahati 781039, Assam, India}

\author{Shawan K. Jha}
\affiliation{Department of Physics, Indian Institute of Technology, Guwahati 781039, Assam, India}

\author{G. A. Bougas}
\affiliation{Department of Physics and LAMOR, Missouri University of Science and Technology, Rolla, MO 65409, USA}

\author{S. I. Mistakidis}
\affiliation{Department of Physics and LAMOR, Missouri University of Science and Technology, Rolla, MO 65409, USA}

\author{Pankaj Kumar Mishra}
\affiliation{Department of Physics, Indian Institute of Technology, Guwahati 781039, Assam, India}

\date{\today}

\begin{abstract}

We investigate Faraday instabilities in three-dimensional parametrically driven trapped dipolar quantum gases within the framework of the
extended Gross-Pitaevskii equation. 
In the superfluid regime, periodic modulation of the short-range interactions induces the resonant excitation of discrete polygonal surface modes displaying sub-harmonic response. 
The resonance frequencies and the parametric windows of instability are accurately captured by an appropriate Mathieu equation, further corroborating our simulations. 
It is also shown that dipolar interactions in superfluids shift the resonance frequencies to lower values compared to their non-dipolar counterparts.
Near the superfluid-to-supersolid transition, intrinsic density undulations associated with the softened roton mode
accelerate pattern formation, yielding hybrid surface and bulk excitations. 
The bulk patterns prevail deeper in the supersolid regime, 
emerging from droplet collisions with the superfluid background. 
Our results reveal a crossover from surface collective modes
to hybrid surface-bulk excitations and demonstrate how parametric driving dictates pattern formation in long-range interacting quantum fluids.

\end{abstract}

\flushbottom

\maketitle
%%%%%%%%%%%%%%%%%%%%%%%%%%%%%%%%%%%%%%%%%%%%%%%%%%%%%%%%%%%%%%%
\section{Introduction}\label{sec:intro}

Originally observed on the surface of classical fluids subject to vertical periodic driving, Faraday patterns constitute a paradigmatic example of spontaneous structure formation in driven nonlinear systems~\cite{Faraday1831,kumar1994parametric,Cross2009,benjamin1954stability,miles1990parametrically,Cross1993}.
These patterns manifest as standing-wave density modulations oscillating at half the driving frequency, and arise from parametric resonances~\cite{GarcaRipoll1999,Kevrekidis2000ParametricQuantumResonances,Sanmartin_parametric_1984}.
The latter occur in systems with periodically driven parameters, and are characterized by the exponential growth of dynamically unstable collective modes~\cite{benjamin1954stability,miles1990parametrically}.
The instability spectrum is determined by the underlying excitation structure of the fluid, rendering pattern formation a sensitive probe of its spectral properties and emergent collective behavior~\cite{vasic2026pattern,kwon2026pattern}. Parametrically driven systems can also exhibit interfacial and surface 
patterns.
In classical fluids, vertically vibrated liquid layers support a wide variety of symmetry-breaking structures, including polygonal and star-shaped patterns emanating from nonlinear coupling among surface modes~\cite{rajchenbach2013observation}.

The realization of Bose-Einstein condensates (BECs) has extended the study of parametric
instabilities into the realm of quantum fluids~\cite{vasic2026pattern,kwon2026pattern}. 
Owing to their exceptional degree of experimental  control,  
these platforms are ideal for investigating nonlinear collective dynamics far from equilibrium~\cite{staliunas2002faraday,engels2007observation,kevrekidis2004pattern,Nath2010,dalfovo1999theory,mistakidis2023few}. In this context, Faraday waves can be generated through periodic modulation of the trapping potential~\cite{Nicolin2007,engels2007observation,GarcaRipoll1999,Staliunas2004,Balaz2012,Hernandez_Faraday_2021} or interaction strength~\cite{staliunas2002faraday,nguyen2019parametric,manna2025pattern,kwon2021spontaneous}, leading to resonant amplification of collective excitations. 
The latter provide a means to explore key properties of BECs, such as the dispersion relation of density waves~\cite{Staliunas2004}, the interfacial tension of immiscible binary BECs~\cite{maity2020parametrically}, and the presence of quantum fluctuations~\cite{nguyen2019parametric}.

In a similar vein, pattern formation and in particular parametric instabilities may be proven powerful tools for unveiling  
microscopic properties of dipolar BECs being characterized by the interplay of short- and long-range interactions~\cite{Chomaz:ROPIP2022,lahaye2009physics,griesmaier2005bose,aikawa2012bose,lu2011strongly,baranov2008theoretical,yi2001trapped,goral2002ground,Bland2022,casotti2024vortices,bottcher2020new}. This competition, together with the presence of quantum fluctuations e.g. in the form of the Lee-Huang-Yang (LHY) energy correction~\cite{lima2011quantum,lima2012beyond}, lead to the appearance of the roton mode in the excitation spectrum~\cite{santos2003roton,o2003rotons,Petter_probing_2019,Hertkorn_density_2021}, which dictates pattern formation.
At the ground state of these settings, the existence of the roton gives rise to spontaneously modulated phases of matter, such as supersolids~\cite{bottcher2019transient,chomaz2019long,tanzi2019observation,Norcia_two_2021} and isolated droplet arrays~\cite{wachtler2016ground,ferrier2016observation,Ferrier_liquid_2016}, both of which have been observed in different experiments utilizing magnetic lanthanide atoms~\cite{Chomaz:2022}.

Pattern formation in non-LHY  dipolar BECs has been thus far investigated in homogeneous superfluid phases following periodic modulation of the contact interactions. 
It was argued that the emergent patterns are characterized by wavenumbers close to the roton minimum~\cite{Nath2010,Lakomy2012,Nadiger2024,Vudragovi2019,Turmanov2020,nicolin_density_2013}.
Accordingly, the excitation of surface and/or bulk modes~\cite{kwon2021spontaneous} in trapped dipolar gases, already within the superfluid region, remain largely unexplored. 
Here, for instance, it is  interesting to study the impact of dipolar interactions, the LHY contribution as well as the trap aspect ratio on the location and behavior of parametric resonances. 
Moreover, it is currently unclear how the presence of intrinsic density modulations can alter the system's response and hence the plausible manifestation of parametric instabilities, potentially involving coupled surface and bulk excitations, under time-periodic driving. 

Motivated by these open questions, we demonstrate the dynamical generation of Faraday patterns in a three-dimensional (3D) harmonically trapped dipolar BEC. 
The latter is initiated  either at the superfluid state or at the superfluid-to-supersolid transition region being subjected to periodic modulation of the scattering length and the external trap respectively. 
Our simulations are based on the appropriate extended Gross–Pitaevskii equation (eGPE)~\cite{Chomaz:ROPIP2022}, revealing
that parametrically driven trapped dipolar BECs feature enriched Faraday instabilities as compared to non-dipolar superfluids~\cite{kwon2021spontaneous,engels2007observation,Zhang2020}. 
We explicate that these instabilities depend crucially
on the trap aspect ratio as well as the relative strength of contact and dipolar interactions.

Specifically, within the dipolar superfluid phase, we find that driving the contact interaction enables to  selectively excite discrete polygonal surface modes characterized by well-defined rotational symmetries and sub-harmonic response. 
The resonance spectrum of the parametric instability is extracted here analytically by deriving the suitable Mathieu equation of a trapped dipolar gas, which takes into account both contact and dipolar interactions in contrast to previous treatments~\cite{Nath2010,Lakomy2012,Nadiger2024,Vudragovi2019,Turmanov2020,nicolin_density_2013}.
The analytical predictions are in adequate agreement with the eGPE simulations. 
It turns out that dipolar interactions shift the resonance frequencies to lower values, while in the limit of strong axial confinement where the contact interaction dominates, we retrieve the resonance spectrum of non-dipolar BECs~\cite{Parker_Thomas_2008}. In sharp contrast, close to the superfluid-to-supersolid transition, we uncover the emergence of coupled surface and bulk modes, a behavior that is traced back to the underlying roton softening~\cite{Recati_supersolidity_2023}. 
Moreover, the small undulations in the ground states at the crossover region favor the faster  onset of pattern formation compared to dipolar superfluids~\cite{bougas2026generation}.
Turning to supersolids, bulk density patterns
consisting of a few droplets prevail, emanating from collisions between the original crystals and the superfluid background. 
Hence, a dynamical crossover takes place from purely surface modes (superfluids) to hybrid excitations (superfluid-to-supersolid transition) and eventually to bulk modes (supersolid regime).

This work is organized as follows. 
In Sec.~\ref{sec:MFmodel}, the 3D eGPE framework describing the dipolar gas  is introduced.  
Section~\ref{sec:GS} discusses the ground-state superfluid and supersolid phases considered in this work, while Sec.~\ref{sec:Protocol} presents the periodic driving protocols employed to generate Faraday patterns.  
Surface pattern formation in superfluids triggered by periodic driving of the contact interactions is analyzed in Sec.~\ref{sec:density_dyn}. 
In Sec.\ref{sec:Mathieu}, the Mathieu description of the surface-mode dynamics is derived, whilst the dependence of the resonant modes on the contact interaction strength and trap aspect ratio is investigated in Sec.~\ref{sec:Mathieu} and  Sec.~\ref{sec:int_depend_reson}.  
The creation of surface and bulk modes induced by periodic driving of the external confinement at the superfluid-to-supersolid crossover and in the supersolid regime is presented in Sec.~\ref{trasitions}. 
Section~\ref{summary and conclusion} offers a summary of our findings and elaborates on future perspectives based on our results. 
In Appendix~\ref{app:scattering}, we demonstrate the equivalence between scattering length and trap periodic modulation in the superfluid regime, while in  Appendix~\ref{app:mathieu} further details on the derivation of the Mathieu equation are outlined. Finally, Appendix~\ref{app:strong} discusses the behavior of the surface modes in the strong transverse confinement limit.

%%%%%%%%%%%%%%%%%%%%%%%%%%%%%%%%%%%%%%%%%%%%%%%%%%%%%%%%%%%%%%%%%%%%%%%%
\begin{figure*}[!htp]
\centering
\includegraphics[width=\textwidth]{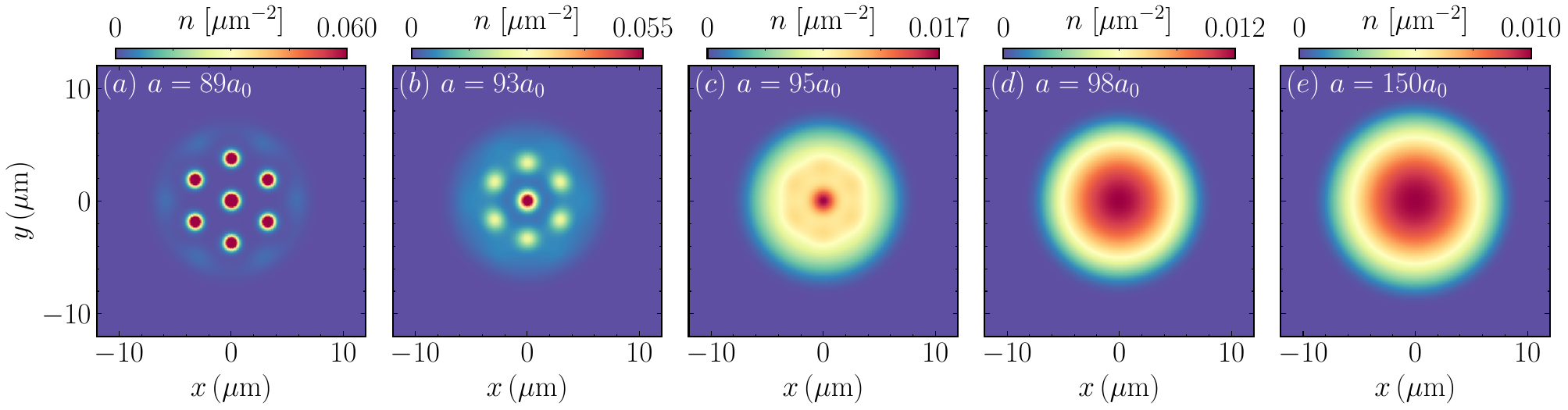}
\caption{Distinct ground state phases of oblate ${}^{164}$Dy dipolar quantum gases at different scattering lengths $a$, as inferred by the
two-dimensional integrated density profiles, $n = \int |\Psi(\boldsymbol{r})|^2 dz$. (a), (b) Supersolid states manifest at small scattering lengths, displaying crystalline order and a superfluid background. (c) As the scattering length increases, the density modulation becomes less pronounced, signaling the supersolid–to-superfluid crossover. (d), (e) When the short-range interactions dominate over the dipolar ones, the gas resides in the trapped dipolar superfluid configuration. Here, $N=8\times 10^4$ and the trapping frequencies are $(\omega_r,\omega_z)= 2\pi \times (43,133)~\rm{Hz}$.
}
\label{fig:Ground_states}
\end{figure*}
%%%%%%%%%%%%%%%%%%%%%%%%%%%%%%%%%%%%%%%%%%%%%%%%%%%%%%%%%%%%%%%%%%%%%%%%

\section{Dipolar quantum gas}
\label{sec:MFmodel}

We consider $N=8\times10^4$ ${}^{164}$Dy atoms, at zero temperature, confined in an oblate harmonic trap with radial and transverse frequencies $(\omega_r,\omega_z) = 2\pi \times (43,133)~\rm{Hz}$, similarly to recent experiments~\cite{Bland2022,casotti2024vortices}. The magnetic atoms (dipoles) are polarized along the $z$ axis by means of an external magnetic field.
The statics and dynamics of the dipolar quantum gas are adequately described by the condensate wave function $\Psi(\mathbf r,t)$, with $\mathbf{r}=(x,y,z)$, whose behavior is governed by the 3D eGPE~\cite{wachtler2016ground,ferrier2016observation,Ronen_dipolar_2006,Bortolotti_scattering_2006,Chomaz:ROPIP2022}
\begin{multline}
i\hbar\, \partial_t \Psi(\mathbf{r}, t) =
- \frac{\hbar^2}{2m} \nabla^2 \Psi
+ V(\mathbf r)\Psi
+ \frac{4\pi\hbar^2 a N}{m}  |\Psi|^2 \Psi \\
+ N\left( \int d\mathbf r' \,
U_{\mathrm{dd}}(\mathbf r-\mathbf r')
|\Psi(\mathbf r',t)|^2 \right)\Psi
+ \gamma(\varepsilon_{\mathrm{dd}}) N^{3/2}|\Psi|^3\Psi .
\label{eq:1}
\end{multline}
Here, $m$ denotes the atomic mass and $V(\mathbf r)= \sum_{k=r,z} m\omega_k^2 k^2/2$ represents the external trapping potential. 
The condensate wave function is normalized according to $\int |\Psi(\mathbf r,t)|^2 d\mathbf r = 1$.

The dipolar atoms feature short-range interactions, modeled by the third term in Eq.~\eqref{eq:1} and characterized by 
the $s$-wave scattering length $a$. The latter can be experimentally tuned using  magnetic Fano–Feshbach resonances~\cite{Inouye1998,Chin2010,Maier_broad_2015}. The fourth term in Eq.~\eqref{eq:1} accounts for the long-range and anisotropic dipole-dipole interaction (DDI)~\cite{yi2001trapped,goral2002ground}. It is described by
\begin{equation}
U_{\mathrm{dd}}(\mathbf r)
=
\frac{3\hbar^2 a_{\mathrm{dd}}}{m}
\frac{1-3\cos^2\theta}{r^3},
\label{Eq:Dipolar_int}
\end{equation}
where $\theta$ is the angle between the relative distance of two dipoles, $\mathbf r=(x,y,z)$, and the polarization direction of the magnetic field. 
The strength of the dipolar interactions is given by the dipolar length, which for $^{164}\mathrm{Dy}$ atoms is $a_{\mathrm{dd}}=131~a_0$ with $a_0$ denoting the Bohr radius.

The last term in Eq.~\eqref{eq:1} represents  the first-order beyond mean-field LHY quantum correction which is valid within the local-density approximation~\cite{lima2011quantum,lima2012beyond}. It accounts for quantum fluctuations  %The LHY term 
and takes the explicit form 
\begin{equation}
\gamma(\varepsilon_{\mathrm{dd}})
=
\frac{128\sqrt{\pi}\hbar^2 a^{5/2}}{3m }
\left(1+\frac{3}{2}\varepsilon_{\mathrm{dd}}^2\right),
\label{Eq:LHY}
\end{equation}
where $\varepsilon_{\mathrm{dd}} \equiv a_{\mathrm{dd}}/a$.
This repulsive contribution counterbalances the attractive dipolar interactions occurring due to head-to-tail dipole collisions~\cite{lahaye2009physics}. 
Depending on the value of $\varepsilon_{\mathrm{dd}}$ determining the interplay between short-range interactions and long-range dipolar forces, the LHY contribution leads to stable ground-state exotic configurations~\cite{Chomaz:ROPIP2022} such as quantum droplet lattices, supersolid and superfluid states some of which are discussed in Sec.~\ref{sec:GS}. 
Accordingly, these states may feature complex nonequilibrium behavior, characteristic of dipolar quantum fluids~\cite{Chomaz:2022}, for instance, when subjected to external periodic driving which is the main focus of our work.

In the following, we employ harmonic oscillator units, i.e. the spatial and temporal scales are measured in terms of the oscillator length $l_r=\sqrt{\hbar / (m\omega_r)}$, and $\omega_r^{-1}$ respectively, while the 3D wave function is expressed with respect to $l_r^{-3/2}$. 
All numerical simulations are performed on a 3D mesh grid discretized by $N_x=N_y=384$ and $N_z=64$ points in the individual spatial directions and featuring periodic boundary conditions. 
We use spatial and temporal discretizations ($dx=dy=0.08l_r, dz=0.2l_r$) and $dt=10^{-4}\omega_r^{-1}$ respectively. 
The ground states of the dipolar gas are obtained using imaginary-time propagation, while the subsequent nonequilibrium dynamics induced by parametric driving [see also Sec.~\ref{sec:Protocol}] is computed through real-time evolution using the time-splitting sine-pseudospectral (TSSP) method~\cite{Rawat2025}. 
The dipolar interaction term is evaluated efficiently in momentum space via Fourier-space convolution with the corresponding dipolar kernel~\cite{kumar2015fortran,goral2002ground}. 

%%%%%%%%%%%%%%%%%%%%%%%%

\section{Ground state phases}
\label{sec:GS}

The ground-state properties of a dipolar quantum gas are governed by the competition between short-range interactions and long-range dipolar forces. 
In particular, by tuning the $s$-wave scattering length $a$, one controls the relative interaction strength ($\varepsilon_{\mathrm{dd}}$), and thereby the equilibrium phase of the system~\cite{Nath2010,bougas2026generation,Chomaz:ROPIP2022}.

%%%%%%%%%%%%%%%%%%%%%%%%%%%%%%%%%%%%%%%%%%%%%%%%%%%%%%%%%%%%%%%%%%%
\begin{figure*}[!htp]
\centering
\includegraphics[width=0.98\textwidth]{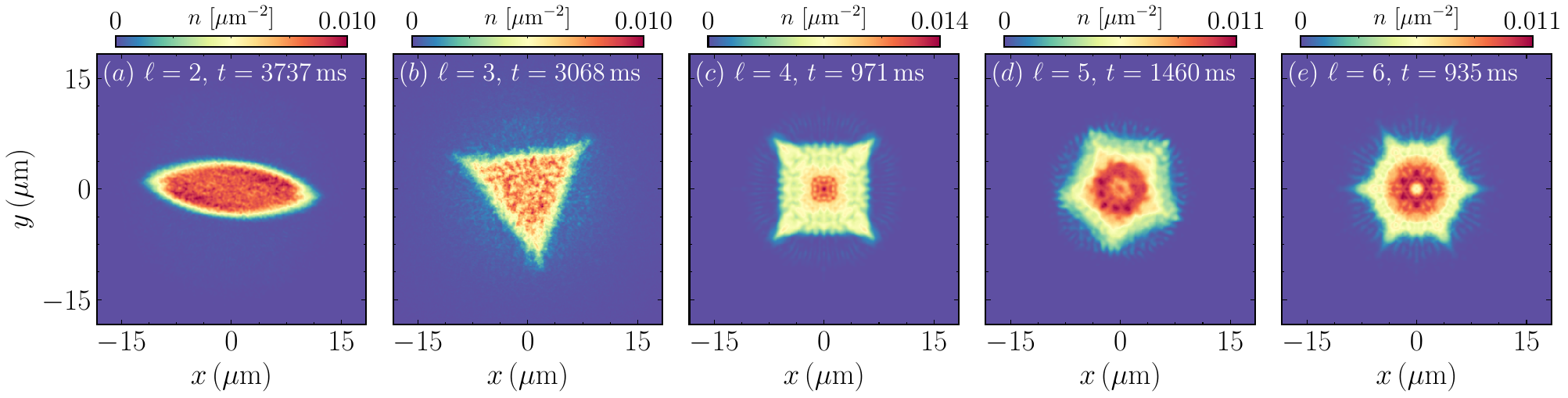}
\caption{Density profiles of different surface patterns of dipolar superfluids arising due to parametric resonances induced by the modulated scattering length [Eq.~(\ref{Eq:Driving})] at specific driving frequencies.
The various surface deformations are classified according to the $\ell$ azimuthal quantum number, and they refer to (a) elliptical ($\ell=2$), (b) triangular ($\ell=3$), (c) square ($\ell=4$), (d) pentagonal ($\ell=5$), and (e) hexagonal ($\ell=6$) surface modes appearing at different time-instants, see also Fig.~\ref{fig:tri_polygon_temporal} and Fig.~\ref{fig:spectral_strength_modes}. These surface modes are excited at driving frequencies $\omega_d/2\pi \simeq 112,\,129,\,147,\,162$, and $174~\rm{Hz}$, respectively. 
The dipolar superfluid is initialized at a scattering length
$a_i = 150~a_0$ and it is subjected to a modulation amplitude $\mathcal{A}=0.25$.
} 
\label{fig:polygon_patterns}
\end{figure*}
%%%%%%%%%%%%%%%%%%%%%%%%%%%%%%%%%%%%%%%%%%%%%%%%%%%%%%%%%%%%%%%%%%%

Figures~\ref{fig:Ground_states}(a)-(e) illustrate two characteristic phases of dipolar gases, namely superfluids and supersolids as well as their crossover regime.
We remark that for scattering lengths $a<87~a_0$, the dipolar gas exhibits the droplet lattice state featuring hexagonal crystalline order and absence of phase coherence~\cite{Norcia_two_2021}.
This state is not shown here since it turns out that upon periodic driving it does not favor pattern formation but rather lattice vibrational modes. 
On the other hand, for scattering lengths $87~a_0 \lesssim a \lesssim 93~a_0$ [Fig.~\ref{fig:Ground_states}(a), (b)], the dipolar interactions dominate over the short-range repulsion, leading to the formation of a supersolid state. 
This is characterized by a hexagonal array of self-organized droplets on top of a superfluid background. In this regime, the system simultaneously exhibits density modulation and global phase coherence, indicating the coexistence of crystalline order and superfluidity~\cite{Bland2022}. As $a$ further increases, i.e. $94~a_0 \lesssim a \lesssim 96~a_0$ [Fig.~\ref{fig:Ground_states}(c)], the droplet structures  gradually disappear, and specifically density modulations become less pronounced. This crossover regime marks the transition from supersolid-to-superfluid states. 
For larger scattering lengths, $a \gtrsim 98~a_0$ [Fig.~\ref{fig:Ground_states}(d), (e)], the repulsive short-range interaction dominates over the dipolar forces. The system, thus, enters the dipolar superfluid phase characterized by a Thomas-Fermi density profile due to the external trap~\cite{Eberlein2005}. 
In this regime, the system becomes less sensitive to $s$-wave scattering length variations (as compared to supersolids), and behaves similarly to weakly interacting trapped non-dipolar BECs.

%%%%%%%%%%%%%%%%%%%%%%%%%%%%%%%%%%%%%%%%%%

\section{Periodic driving protocols}
\label{sec:Protocol}

In order to dynamically excite Faraday patterns in dipolar gases, we first focus on the dipolar superfluid regime [Fig.~\ref{fig:Ground_states}(e)].
This offers a well-established connection with non-dipolar repulsively interacting BECs, which have been proven reliable platforms to realize and accommodate Faraday patterns following periodic modulations of e.g. the interparticle interactions~\cite{kwon2021spontaneous,Staliunas2002,nguyen2019parametric}. 
Therefore, in direct analogy with non-dipolar BECs, we consider parametric driving 
of the scattering length according to
\begin{equation}
a(t)
=a_i
\left[1+\mathcal{A}\cos(\omega_d t)\right].
\label{Eq:Driving}
\end{equation}
Here, $a_i$ denotes the initial (at $t=0$) scattering length, $\mathcal{A}$ is the dimensionless modulation amplitude, and $\omega_d$ is the driving frequency. 
Evidently, the total modulation amplitude of the scattering length is given by $a_m = a_i \mathcal{A}$.
Unless stated otherwise, in what follows, we consider $a_i=150~a_0$, which corresponds to a dipolar superfluid state, $\mathcal{A}=0.25$, and hence $a_m =37.5~a_0$ ensuring that the system  remains within the superfluid regime in the course of the driving. 
Moreover, the driving frequency, $\omega_d$, is varied to identify distinct parametric resonances.

In contrast, the time-periodic modulation of the scattering length is not a suitable protocol to produce Faraday patterns in the case of supersolids and in the superfluid-to-supersolid crossover. 
Especially in the crossover region the application of this protocol facilitates the dynamical crossing of the phase transition which can yield additional excitations associated with complex response that may even lead to a turbulent cascade~\cite{bougas2026generation}. 
Overall, in both of these regimes, it turns out that the underlying density distributions change  significantly even upon small scattering length variations, in contrast to dipolar superfluids, cf. Fig.~\ref{fig:Ground_states}(b), (c) and Fig.~\ref{fig:Ground_states}(d), (e). 
To circumvent this issue, we instead consider periodic modulations of the radial trap frequency in the case of supersolids, according to 
\begin{equation}
\omega_r^2(t) = \omega_r^2(0) \left[    1 + \mathcal{A} \cos(\omega_d t)   \right],    
\label{Eq:Driving_trap}
\end{equation}
where $\omega_r \equiv \omega_r(0) = 2\pi \times 43~\rm{Hz}$ refers to the radial trap frequency. Also, as before $\mathcal{A}$ represents the dimensionless modulation amplitude, while $\omega_d$ is the driving frequency.
It is interesting to note that the confinement and interaction modulation protocols outlined in Eq.~\eqref{Eq:Driving} and Eq.~\eqref{Eq:Driving_trap} respectively excite the same parametric resonances at the same resonant frequencies for dipolar superfluids, as we showcase in Appendix~\ref{app:scattering}.
The same holds for non-dipolar BECs as it was demonstrated in Ref.~\cite{maity2020parametrically}.

\section{Faraday instability of  dipolar superfluids}
\label{sec:SF}

%%%%%%%%%%%%%%%%%%%%%%%%%%%%%%%%%%%%%%%%%%%%%%%%%%%%%%%%%%%%%%%%%%%%%%%%%%%%%
\begin{figure*}[!htp]
\centering
\includegraphics[width=\textwidth]{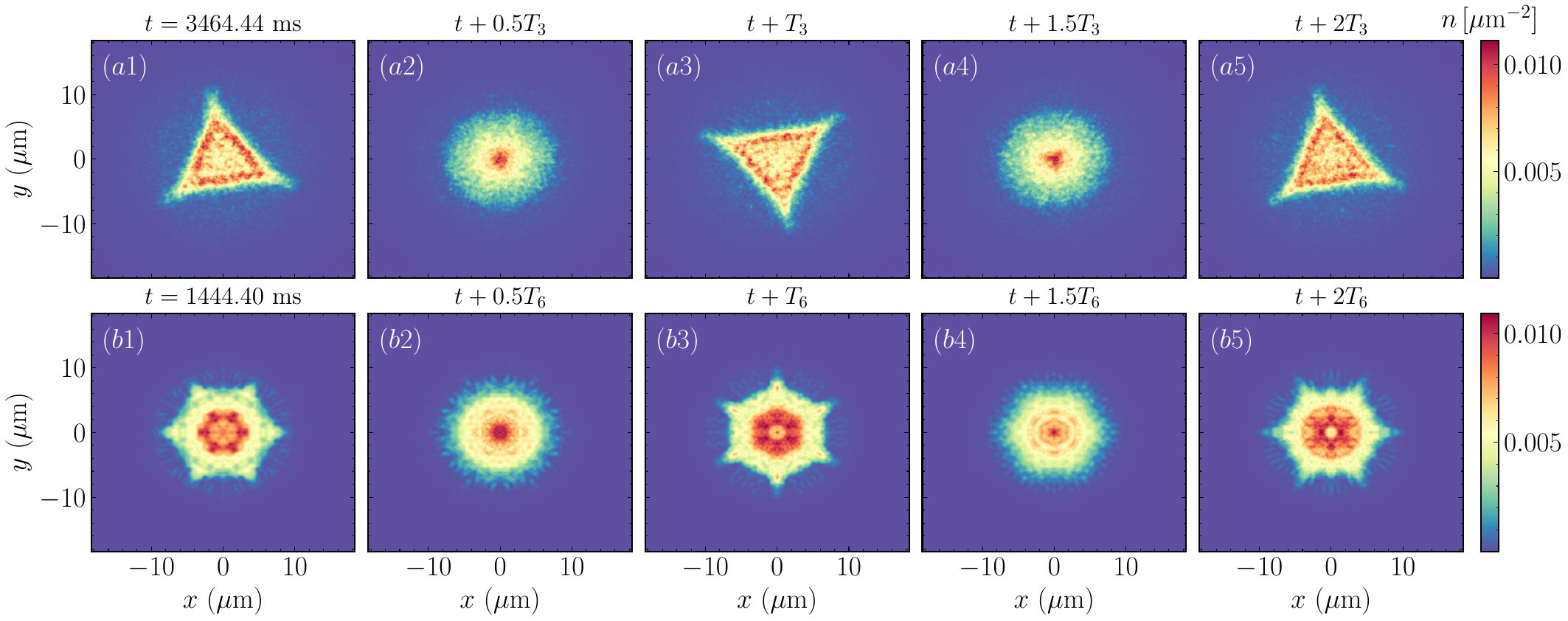}
\caption{Precession, deformation and recurrence of resonantly excited polygonal surface patterns induced by modulation of the scattering length in dipolar superfluids.
(a1)-(a5) [(b1)-(b5)] Density snapshots (see legends) of the dipolar superfluid accommodating a triangular [hexagonal] surface pattern excited at driving frequency $\omega_d/2\pi \simeq 129~\mathrm{Hz}$
[$\omega_d/2\pi \simeq 174~\mathrm{Hz}$].
The patterns exhibit a sub-harmonic response and re-emerge after two driving periods, i.e. $2T_{\ell}$ with $\ell=3,6$, a behavior that is characteristic of first-order parametric resonances. 
All other parameters are the same as in Fig.~\ref{fig:polygon_patterns}.
}
\label{fig:tri_polygon_temporal}
\end{figure*}
%%%%%%%%%%%%%%%%%%%%%%%%%%%%%%%%%%%%%%%%%%%%%%%%%%%%%%%%%%%%%%%%%%%%%%%%%%%%%%

\subsection{Surface density modes}\label{sec:density_dyn}

Dipolar superfluids can accommodate a plethora of surface density patterns when subjected to parametric driving. 
An overview of such surface modes developing on the relevant integrated densities, $n(x,y,t) = \int dz~ \abs{\Psi(\boldsymbol{r},t)}^2$, of the dipolar gas is provided in  Fig.~\ref{fig:polygon_patterns} following scattering length modulation given by Eq.~(\ref{Eq:Driving}) at different driving frequencies and fixed amplitude $\mathcal{A}$. 
More concretely, the arising patterns exhibit distinct azimuthal symmetries, such as triangular [Fig.~\ref{fig:polygon_patterns}(b)], square [Fig.~\ref{fig:polygon_patterns}(c)], or hexagonal [Fig.~\ref{fig:polygon_patterns}(e)] configurations.
Hence, the azimuthal number $\ell$ becomes a good quantum number for describing the emergent configurations.
Moreover, the number of vertices of the polygonic patterns, i.e. the $\ell$ number, can be systematically increased for larger  driving frequency.
For instance, surface patterns corresponding to the $\ell = 2, 3, 4, 5,$ and $6$ modes are observed at driving frequencies $\omega_d/2\pi \simeq 112$, 129, 147, 162, and 174~Hz, respectively.
As it will be demonstrated in the following [see also Sec.~\ref{sec:Mathieu}], these structures emanate from the parametric excitation of unstable surface modes activated at specific driving frequency intervals whose location is interaction dependent, see also the discussion in Sec.~\ref{sec:int_depend_reson} and Fig.~\ref{fig:fd_vs_scattering_and_sigma}(b).

All these unstable modes occur at relatively late evolution times, $t\gtrsim 900~\rm{ms}$. However, the instability onset, and consequently the appearance of the polygonal patterns, can be significantly accelerated by introducing a weak random Gaussian perturbation to the initial state, an effect that is naturally present in experiments. Indeed, by considering a perturbed initial state, $\Psi(\mathbf{r},0)=\Psi_0(\mathbf{r})[1+\epsilon\,\eta(x,y)]$, where $\Psi_0$ is the ground-state wavefunction, $\eta(x,y)$ is a Gaussian random distribution with zero mean and unit variance, and $\epsilon=10^{-3}$ is the perturbation  amplitude, we observe the expedition of the pattern formation.
For instance, the $\ell=2$, $\ell=4$, and $\ell=6$ surface modes emerge at approximately $1500~\mathrm{ms}$, $600~\mathrm{ms}$, and $510~\mathrm{ms}$, respectively.
Recall that such surface modes have already been observed in non-dipolar superfluids~\cite{kwon2021spontaneous}. However, as we will argue below [Sec.~\ref{sec:Mathieu} and Fig.~\ref{tongue}] the parametric resonances are shifted in the presence of dipolar interactions even  within the superfluid regime.

Focusing first on the triangular configuration ($\ell =3$) whose dynamical density evolution is presented in Fig.~\ref{fig:tri_polygon_temporal}(a1)-(a5), we observe that it features a clear recurrent behavior. 
Starting from a reference time-instant $t$, at which the dipolar gas exhibits a triangular shape [Fig.~\ref{fig:tri_polygon_temporal}(a1)], the latter subsequently rotates by a $\pi/3$ angle at a time interval  $t+T_3$ [Fig.~\ref{fig:tri_polygon_temporal}(a3)].
The $T_3$ timescale coincides with one period of the modulated scattering length, i.e. $T_3= 2\pi / \omega_d$.
After two driving periods, $t+2T_3$, the triangular pattern completes a full dynamical cycle [Fig.~\ref{fig:tri_polygon_temporal}(a5)] and its spatial structure closely resembles the one at the reference time $t$.
This confirms that the triangular pattern exhibits a sub-harmonic response with a period $2T_3$, in agreement with the characteristic signature of first-order parametric resonance [see also Sec.~\ref{sec:Mathieu}].
At half driving periods, i.e. $t+0.5T_3$ or $t+1.5T_3$, the triangular surface pattern disappears, see Fig.~\ref{fig:tri_polygon_temporal}(a2), (a4). Instead, during these time instants the dipolar superfluid mostly resembles the ground state configuration at the superfluid-to-supersolid crossover [see Fig.~\ref{fig:Ground_states}(c)]. 
Note that the choice of the reference time $t$ in Fig.~\ref{fig:Ground_states}(a1) does not affect the subsequent periodic evolution. Indeed, the same phenomenology persists if we consider other reference time snapshots, e.g. $t+2T_3$.

A similar dynamical response occurs also  for the hexagonal ($\ell = 6$) mode, as shown in Fig.~\ref{fig:tri_polygon_temporal}(b1)-(b5). 
The six-lobed surface pattern develops when the scattering length is driven with a frequency $\omega_d/(2\pi) = 174~\rm{Hz}$, see e.g. the density at the $t=1444.40~\rm{ms}$ reference time instant in Fig.~\ref{fig:tri_polygon_temporal}(b1).
Due to the six-fold rotational symmetry of the mode, the pattern rotates by a $\pi/6$ angle after one driving period, i.e. $t+T_6$ [Fig.~\ref{fig:tri_polygon_temporal}(b3)], where $T_6 = 2\pi/\omega_d$.
Similarly to the $\ell=3$ mode, the hexagonal surface pattern precesses back to its original configuration at the reference time $t$ after two driving periods, $t+2T_6$ [Fig.~\ref{fig:tri_polygon_temporal}(b5)].
Furthermore, at half driving periods [Fig.~\ref{fig:tri_polygon_temporal}(b2), (b4)], the hexagonal patterns are destabilized, and the original rotational symmetry is approximately restored, albeit the dipolar superfluid is significantly excited. 
Such a sub-harmonic response takes place for all patterns depicted in Fig.~\ref{fig:polygon_patterns}.

To quantify the emergence of different polygonal surface patterns, we analyze the time-averaged spectral strength $\langle F_\ell(t) \rangle$, which characterizes the amplitude of the $\ell$-th azimuthal deformation mode. For this purpose, we first extract the condensate boundary $R(\phi,t)$ as a function of the polar angle $\phi$, measured from the center of mass of the dipolar gas~\cite{kwon2021spontaneous}. 
The spectral strength of the $\ell$-th mode is then obtained as the normalized magnitude of the corresponding Fourier coefficient,
\begin{equation}
F_\ell(t) = \frac{1}{N_\phi} \left| \sum_{j=0}^{N_\phi-1} R(\phi_j,t)\, e^{-i \ell \phi_j} \right|,
\end{equation}
where $N_\phi$ is the number of grid points in the angular direction. 
This observable provides a direct measure of the strength of the $\ell$-fold symmetric deformation of the surface and it is experimentally tractable via in-situ absorption imaging as it was done in Ref.~\cite{kwon2021spontaneous}.
Subsequently, $F_{\ell}(t)$ is monitored stroboscopically by applying a moving-time average over one driving period, $2\pi / \omega_d$, yielding $\braket{F_{\ell}(t)}$.

\begin{figure}[t]
\centering
\includegraphics[width=\columnwidth]{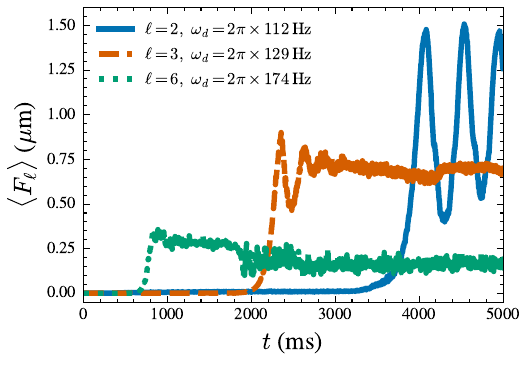}
\caption{Time-averaged spectral strength, $\langle F_{\ell}(t) \rangle$, associated with surface deformation modes of different angular symmetries (see also main text). 
The higher $\ell$ modes manifest faster as compared to the lower ones (see legends), whereas the latter are characterized by stronger surface deformations, as captured by their larger  spectral strengths. 
}
\label{fig:spectral_strength_modes}
\end{figure}

The temporal evolution of the time-averaged spectral strength exhibits three distinct dynamical stages, see Fig.~\ref{fig:spectral_strength_modes}. 
At early times, before the instability onset, $\braket{F_{\ell}(t)} =0$ for all modes. In this time-interval, the dipolar superfluid retains its radial symmetry while expanding and contracting, i.e. undergoing a collective breathing mode (not shown for brevity), while the surface patterns have not yet been developed. 
Consecutively, the spectral strength increases rapidly, a behavior that is consistent with the exponential amplification of unstable modes occurring in parametric resonances~\cite{liebster2025observation}.
The time-instant and the amplitude of this growth stage of $\braket{F_{\ell}(t)}$ depend strongly on the activated surface modes. It is found that the higher the $\ell$-fold symmetry the faster the parametric instability arises, a property that has been observed in non-dipolar superfluids~\cite{kwon2021spontaneous}.
For instance, the hexagonal mode ($\ell=6$) manifests at $t \sim 800~\rm{ms}$, whereas the elliptical mode ($\ell=2$) appears much later, $t \sim 3800 ~ \rm{ms}$ [Fig.~\ref{fig:spectral_strength_modes}].
In the third dynamical stage, the spectral strength oscillates around an approximately constant value. 
The latter denotes the strength of the azimuthal deformations associated with the $\ell$-mode.
As expected, the lower $\ell$ surface patterns are characterized by a larger spectral strength. 
This is corroborated by the more pronounced surface deformations of the lower $\ell$-mode  integrated densities [Fig.~\ref{fig:polygon_patterns}] as compared to higher ones.

%%%%%%%%%%%%%%%%%%%%%%%%%%%%%%%%%%%%%%
%%%%%%%%%%%%%%%%%%%%%%%%%%%%%%%%%%%%%%%%%%%%%%%%%%%%%%%%

%%%%%%%%%%%%%%%%%%%%%%%%%%%%%%%%%%%%%%%%%%%%%%%%%%%%%%%%%
\subsection{Mathieu analysis of surface-mode instability}
\label{sec:Mathieu}
 
To understand the appearance of distinct polygonal patterns at specific driving frequencies, and the parametric instabilities responsible for them, we derive an equation of motion for small deformations of the condensate surface. The starting point is the Madelung transformation~\cite{madelung1927quantum},
\begin{equation}
\Psi(\mathbf{r}, t) = \sqrt{\frac{\rho(\mathbf{r}, t)}{N}}\, e^{i \varphi(\mathbf{r}, t)},
\end{equation}
where $\rho(\mathbf{r},t)$ is the three-dimensional density and $\varphi(\mathbf{r},t)$ the phase of the wavefunction. Inserting this ansatz back into the eGPE [Eq.~(\ref{eq:1})] yields the hydrodynamic equations~\cite{dalfovo1999theory},
\begin{subequations}
\begin{equation}
\frac{\partial \rho}{\partial t}
+
\boldsymbol{\nabla}\!\cdot(\rho\mathbf v)
=
0,
\end{equation}
\begin{equation}
m\frac{\partial\mathbf v}{\partial t}
=
-\boldsymbol{\nabla}
\left[
\frac12 m\mathbf{v}^2
+
V(\mathbf r)
+
g(t)\rho
+
\Phi_{\rm{dd}}[\rho]
\right],
\label{eq:velocity derivative}
\end{equation}
\end{subequations}
where $\mathbf{v}= \hbar/m \boldsymbol{\nabla} \varphi$ is the hydrodynamic velocity, and $g(t)=4\pi \hbar^2  a(t)/m$.
%Here $g(t)$ is the (possibly time-modulated) contact coupling, and
Moreover, $\Phi_{\rm{dd}}[\rho] = \int d\mathbf{r}'~ U_{\rm{dd}}(\mathbf{r}-\mathbf{r}') \rho(\mathbf{r}')$ is the dipolar interaction potential.
The LHY term has been neglected in Eq.~\eqref{eq:velocity derivative}, since its contribution is negligible in dipolar superfluids.
Certainly, incorporating beyond mean-field effects, and deriving a generalized hydrodynamic framework along the lines of Ref.~\cite{Poli_excitations_2024} is a worth pursuing future direction. 
Such a formulation can be combined with Floquet analysis to explore pattern formation in the supersoid regime.
Since we will focus on the early time instants preceding pattern formation, density deformations are small, and hence the kinetic term (quantum pressure) has been also omitted.

We consider small density deformations around the equilibrium density, $\rho_0$, namely $\rho=\rho_0 + \delta \rho$, as well as small superfluid velocity fluctuations, $\mathbf{v}= \delta \mathbf{v}$. 
The equilibrium density, $\rho_0$, is approximated to a good degree by a Thomas-Fermi profile for dipolar gases~\cite{odell2004exact,Eberlein2005}, which provides an analytically tractable handle on the parametric resonances. 
Specifically, linearizing the hydrodynamic equations leads to the following wave equation for the density perturbations,
\begin{equation}
m
\frac{\partial^2\delta\rho}{\partial t^2}
=
\boldsymbol{\nabla}\!\cdot\!
\left[
\rho_0
\boldsymbol{\nabla}
\Big(
g(t)\,\delta\rho
+
\Phi_{\rm{dd}}[\delta\rho]
\Big)
\right].
\label{eq:linear_density_wave}
\end{equation}
To account for the emerging polygonal $\ell$-patterns, we assume deformations along the planar direction of the form
\begin{equation}
\delta\rho(r,\phi,t)
=
\zeta_\ell(t)\,r^\ell\cos(\ell\phi).
\label{Eq:mode_expansion}
\end{equation}
Here, $(r,\phi)$ are the polar coordinates, and $\zeta_\ell(t)$ designates the relatively small deformation  amplitude. Substituting this ansatz into Eq.~(\ref{eq:linear_density_wave}),  yields a Mathieu equation for $\zeta_\ell(t)$ [for further details see Appendix~\ref{app:mathieu}], which reads
\begin{subequations}
\begin{gather}
\frac{d^2\zeta_\ell}{dt^2}
+
\omega_\ell^2
\left[
1+b\cos(\omega_d t)
\right]
\zeta_\ell
=0,
\label{eq:mathieu_final} \\
\omega_{\ell}^2
=
\ell\,\omega_r^2
+
\ell\,\frac{2A_\rho}{m}
+
\ell\,\frac{2A_\ell}{m},
\label{eq:omega} \\
b=\frac{\ell\left(\omega_r^2+\dfrac{2A_\rho}{m}\right)}
{\omega_\ell^2}
\,\mathcal{A} \ll 1.
\end{gather}
\end{subequations}

It becomes evident that the natural frequencies, $\omega_{\ell}$, [Eq.~(\ref{eq:omega})] are composed of three contributions. The first term, $\ell \omega_r^2$ is identical with the natural frequency appearing in a non-dipolar superfluid~\cite{kwon2021spontaneous}. 
The second term, $A_{\rho}$, stems from the dipolar interactions, $\Phi_{\rm{dd}}[\rho_0]$, in the equilibrium configuration. On the other hand $A_{\ell}$ emanates from the perturbed dipolar potential, $\Phi_{\rm{dd}}[\delta \rho]$, caused by the density deformations of Eq.~\eqref{Eq:mode_expansion}. 
Eventually it can be shown that the coefficients $A_{\rho}$ and $A_{\ell}$ are fully determined by the equilibrium configuration of the dipolar superfluid. In particular, they are expressed directly in terms of the Thomas-Fermi radii in the planar and axial directions, see also Appendix~\ref{app:mathieu} for details.
Moreover, $A_{\ell}$ carries an explicit dependence on the azimuthal quantum number. Finally, the parameter $b$ represents the overall driving amplitude.

%%%%%%%%%%%%%%%%%%%%%%%%%%%%%%%%%%%%%%%%%%%%%%%%%%%%%%%%%%
\begin{figure}[!htbp]
\centering
\includegraphics[width=\linewidth]{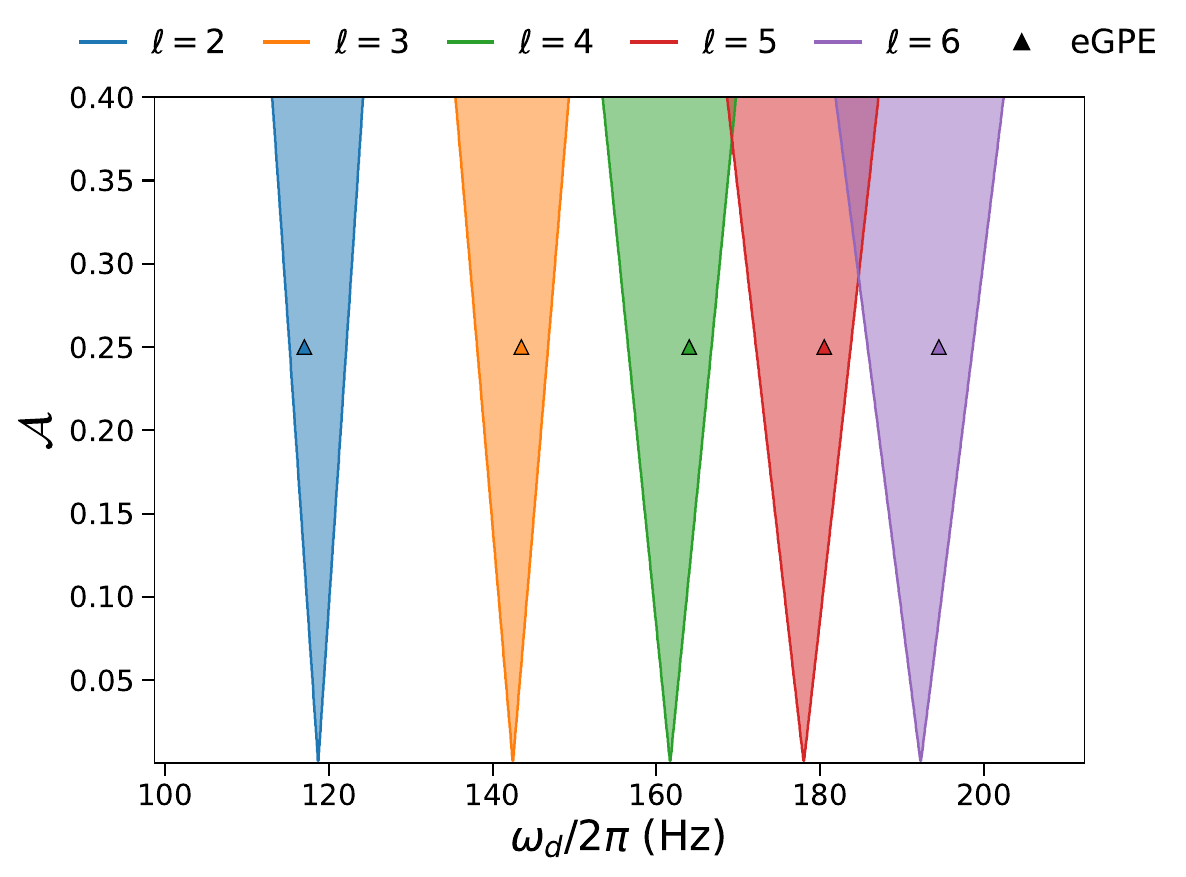}
\caption{Instability tongues of the dipolar superfluid for different $\ell$ modes across the $\mathcal{A}$--$\omega_d$ plane, at $\omega_z=2\pi\times500$~Hz. The resonant frequencies in the small driving amplitude limit, i.e. $\mathcal{A}\to 0$, as obtained analytically (numerically) from the Mathieu equation (eGPE) correspond to $\omega_d/2\pi=118.68$~Hz ($\omega_d/2\pi\simeq117.00$~Hz) for $\ell=2$, $\omega_d/2\pi=143.50$~Hz ($\omega_d/2\pi\simeq141.50$~Hz) for $\ell=3$, $\omega_d/2\pi=161.68$~Hz ($\omega_d/2\pi\simeq164.00$~Hz) for $\ell=4$, $\omega_d/2\pi=178.00$~Hz ($\omega_d/2\pi\simeq180.50$~Hz) for $\ell=5$, and $\omega_d/2\pi=192.28$~Hz ($\omega_d/2\pi\simeq194.50$~Hz) for $\ell=6$. For direct comparison, the resonant frequencies from the eGPE at $\mathcal{A}=0.25$ are also depicted with triangles on top of the Floquet tongues.
}

\label{tongue}
\end{figure}
%%%%%%%%%%%%%%%%%%%%%%%%%%%%%%%%%%%%%%%%%%%%%%%%%%%%%%%%%%%%%%%%%

The Mathieu equation is well suited for determining the parametric regions where small surface deformations grow exponentially, leading eventually to the formation of the observed polygonal structures~\cite{acar2016floquet}, e.g. illustrated in Fig.~\ref{fig:polygon_patterns}. 
To obtain the stability diagram, Eq.~(\ref{eq:mathieu_final}) is integrated over one driving period and the corresponding Floquet multipliers are computed. Parameter ($\mathcal{A},\omega_d$) combinations for which at least one Floquet multiplier has modulus larger than unity are classified as unstable, yielding the Arnold tongue structure (stability diagram) shown in Fig.~\ref{tongue}. 
As it can be seen, the unstable domains form the characteristic Arnold tongues~\cite{Pikovsky_synchronization_2001} of the Mathieu equation in the plane of the modulation strength $\mathcal{A}$ and the driving frequency $\omega_d/2\pi$. 
Each instability tongue corresponds to the resonant excitation of a surface mode characterized by the azimuthal quantum number $\ell$. 
Higher-order surface modes emerge at progressively larger driving frequencies, producing a sequence of instability regions shifted toward higher $\omega_d$~\cite{benjamin1954stability}.
At vanishing driving amplitude, i.e. $\mathcal{A} \to 0$, the $n$-th instability tongue of each $\ell$-mode manifests at roughly $\omega_d \approx 2\omega_{\ell}/n$. 
Increasing the modulation amplitude leads to broadening of the instability tongues, indicating the presence of wider instability regions. 
Outside the unstable domains, the dipolar gas remains dynamically stable exhibiting only small-amplitude expansion and contraction dynamics without developing angular surface deformations. 
This has been further confirmed through eGPE simulations using driving frequencies outside the instability tongues.

The Mathieu framework yields accurate predictions for the resonant  frequencies of patterns characterized by small azimuthal quantum numbers
at tighter transverse confinements. Indeed, considering $\omega_z/(2\pi) = 500~\rm{Hz}$ the agreement between the frequencies stemming from the eGPE [triangles in Fig.~\ref{tongue}] and the Mathieu framework [shaded regions in Fig.~\ref{tongue}] is excellent, as inferred for instance from the stability diagram in Fig.~\ref{tongue}.
A source of discrepancy for the frequencies of high-$\ell$ modes arises from neglecting the quantum pressure term in Eq.~\eqref{eq:velocity derivative}. This contribution 
becomes progressively significant for higher $\ell$-fold patterns, due to their prominent spatial distribution~\cite{stringari1996collective}.
In contrast, for weaker axial confinements the deviations between the two methods become more noticeable.
Specifically, in the case of $\omega_z/(2\pi) = 133~\rm{Hz}$, for $\ell=2$, and $\ell=4$, the resonant driving frequencies stemming from Eq.~\eqref{eq:omega} [the eGPE]  are $\omega_d/(2\pi) = 111.5~\rm{Hz}$ [$112~\rm{Hz}$] and $\omega_d/(2\pi)= 139.58~\rm{Hz}$ [$147~\rm{Hz}$] respectively.
For higher modes an increasing discrepancy is observed; for $\ell=6$ the Mathieu analysis yields $\omega_d/(2\pi) = 156.74~\rm{Hz}$, whereas the eGPE predicts that $\omega_d/(2\pi) = 174~\rm{Hz}$. 
This deviation is attributed, at least in part, 
to the presence of axial excitations [not shown for brevity] which are not taken into account within our hydrodynamic description.
These deviations become more  pronounced for driving frequencies comparable and larger than the axial trapping frequency, $\omega_z$, i.e. for higher $\ell$ modes.

Such a behavior demonstrates that the Mathieu framework, despite the various approximations used, is able to accurately capture the mechanism underlying the parametric excitation of the surface modes.
The Mathieu analysis also naturally explains the spatiotemporal symmetry of the emergent patterns observed in Fig.~\ref{fig:polygon_patterns}. Since the coefficients of the Mathieu equation are periodic in time with period $T=2\pi/\omega_d$, Floquet theory implies that, in the principal parametric instability tongue, the unstable mode is subharmonic with respect to the drive, taking the simple Floquet form $\zeta_\ell(t)\simeq e^{\alpha t}\cos(\omega_d t/2+\chi)$, with $\alpha$ the (small) growth rate and $\chi$ a fixed phase~\cite{kovacic2018mathieu}. Consequently, apart from the slowly growing Floquet envelope, the mode amplitude changes sign after one driving period, $\zeta_\ell(t+T)\simeq-\zeta_\ell(t)$. This sign change is compensated by the azimuthal phase factor under a rotation $\phi\rightarrow\phi+\pi/\ell$, since $e^{i\ell(\phi+\pi/\ell)}=-e^{i\ell\phi}$. Hence the surface deformation is invariant under the combined transformation $(\phi,t)\rightarrow(\phi+n\pi/\ell,t+nT)$, $n\in \mathbb{Z}$, which accounts for the discrete rotational symmetry of the observed polygonal patterns.

%%%%%%%%%%%%%%%%%%%%%%%%%%%%%%%%%%%%%

\subsection{Tuning parametric resonances via interactions and axial trap frequency}\label{sec:int_depend_reson}

To assess the impact of dipole-dipole forces on the surface pattern generation, we consider the dynamical response of the driven dipolar superfluid using the eGPE  with respect to different axial confinements, $\omega_z$.
When the characteristic transverse energy, $\hbar\omega_z$ greatly exceeds the chemical potential of the dipolar gas, $\mu$, namely it holds that $ \hbar \omega_z \gg \mu$, the dipoles become effectively kinematically constrained on the plane.
In the limit of extremely strong axial confinement the nonlocal contribution of the dipolar interactions is strongly suppressed, and the short-range interactions dominate as it was argued in Refs.~\cite{Parker_Thomas_2008,zhen_breaking_2025}.
Accordingly, in this regime the dipolar superfluids behave approximately as normal BECs.
Therefore, the axial confinement is a direct and experimentally accessible probe of the effect of dipolar interactions on the resonance frequencies of the Faraday patterns.

\begin{figure}[!htp]
\centering
\includegraphics[width=\linewidth]{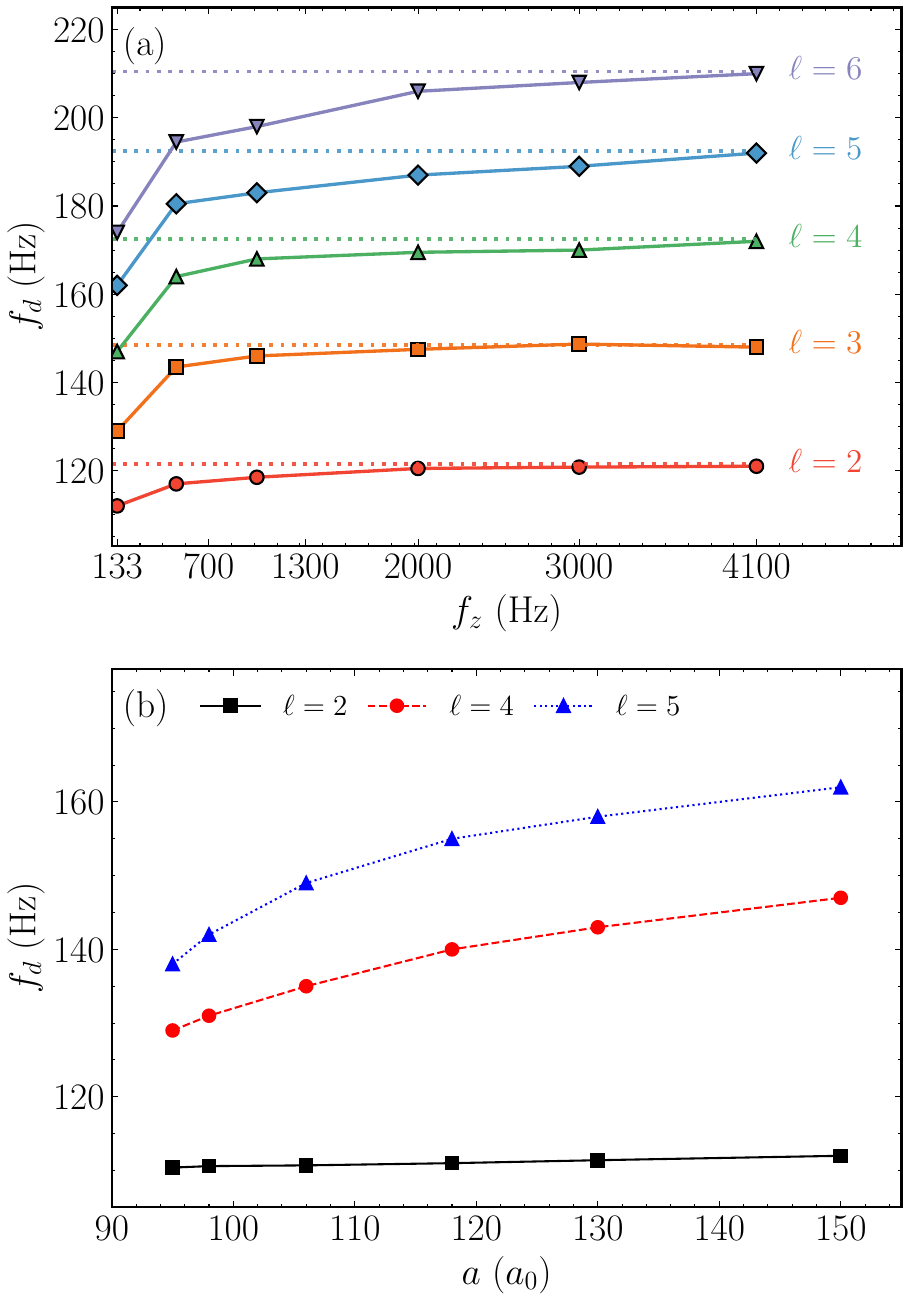}
\caption{
(a) Dependence of the resonant frequencies $f_d$ on (a) the axial trap frequency, $f_z$, and (b) the scattering length $a$ for different polygonal surface modes (see legends) using eGPE simulations. 
The dashed lines for distinct $l$ values in panel (a) mark the resonance frequencies in the absence of dipolar interactions. 
It turns out that the dipolar interaction shifts $f_d$ to smaller values and for strong axial confinement, where the dipolar force is suppressed, they attain their hydrodynamic value in the case of contact interactions. 
A monotonic increase of $f_d$ for larger $a$ occurs [panel (b)] especially for increasing $\ell$. 
In all cases, the driving amplitude is $\mathcal{A}=0.25$.} 
\label{fig:fd_vs_scattering_and_sigma}
\end{figure}

To explicate this effect we monitor the behavior of the resonant driving frequencies, $f_d=\omega_d/(2\pi)$, upon varying $\omega_z$, while keeping all other system  parameters fixed [Fig.~\ref{fig:fd_vs_scattering_and_sigma}(a)]. 
For all considered $\ell=2\text{--}6$ modes, the resonant frequencies shift systematically toward higher values with increasing $\omega_z$. 
This is attributed to the combined behavior of the dipolar contributions, $A_{\rho}$ and $A_{\ell}$, in Eq.~\eqref{eq:omega}, and in particular to their dependence on the ratio of Thomas-Fermi radii which changes with $\omega_z$ [see also Appendix~\ref{app:mathieu} for further details].
Such an increasing trend is especially pronounced for higher $\ell$ surface modes and it can be explained from the natural frequencies, $\omega_{\ell}$, stemming from the Mathieu analysis. 
Specifically, the combined contribution of $A_{\rho}$ and $A_{\ell}$ in Eq.~\eqref{eq:omega} is more pronounced for higher $\ell$, due to the explicit $\ell$-dependence of the $A_{\ell}$ dipolar term [see also Appendix~\ref{app:mathieu}]. The latter stems from the dipolar potential sourced by the density deformations, and thus is highly susceptible to the distinct $\ell$-fold patterns.
As the confinement becomes even stronger, the resonant frequencies gradually approach a saturation regime. 
In this limit, the response of the dipolar gas is akin to that of a non-dipolar superfluid, and we recover the well-known behavior $\omega_{\ell}=\sqrt{\ell} \omega_r$ [marked by the horizontal dotted lines in Fig.~\ref{fig:fd_vs_scattering_and_sigma}(a)], see also Appendix~\ref{app:strong} for further details. 
It can be readily seen that the saturation of the natural frequencies to the hydrodynamic non-dipolar BEC prediction  occurs faster for lower lying surface modes.

To further probe the effect of dipolar interactions on the resultant surface pattern formation, we next investigate the dependence of the resonance frequencies on the scattering length.
Tuning $a$ to values smaller than $150~a_0$, but still within the dipolar superfluid regime, leads to a monotonic decrease of $f_d$ for the surface modes, see Fig.~\ref{fig:fd_vs_scattering_and_sigma}(b) for $\ell=2,4,5$.
This is due to the enhanced role of dipolar interactions, which lead to a systematic reduction of the resonance frequencies, as has been also discussed above and shown in Fig.~\ref{fig:fd_vs_scattering_and_sigma}(a).

%%%%%%%%%%%%%%%%%%%%%%%%%%%%%%%%%%%%%%%%%%%%%%%%%%%%%%%%%%%%%%%%
%%%%%%%%%%%%%%%%%%%%%%%%%%%%%%%%%%%%%%%%%%%%%%%%%%%%%%%%%%%%%%%%%%

Alternatively, this behavior can be understood directly from the Mathieu equation. As the scattering length is reduced, the planar and axial Thomas-Fermi radii contract such that their ratio decreases [see Appendix~\ref{app:mathieu}] compared to its value deep in the superfluid regime ($a=150\,a_0$). Consequently, both dipolar contributions to the natural frequency, $A_\rho$ and $A_\ell$ [Eq.~\eqref{eq:omega}], are modified. 
Such a modification is more pronounced for high-$\ell$ due to the explicit $\ell$-dependence of the $A_{\ell}$ dipolar term [see Appendix~\ref{app:mathieu}].
This feature is clearly revealed in the trend observed in Fig.~\ref{fig:fd_vs_scattering_and_sigma}(b); the $\ell=2$ mode is only weakly affected by changes in $a$, whereas the $\ell=4$ and $\ell=5$ modes exhibit a much stronger dependence. 
Thus in dipolar superfluids the scattering length acts as an effective, mode-dependent control parameter for tuning the parametric response and stability of Faraday surface modes.

%%%%%%%%%%%%%%%%%%%%%%%%%%%%%%%%%%%%%%%%%%%%%%%%%%%%%%%%%%%%%%%%%%

\section{Pattern formation near the supersolid transition}
\label{trasitions}

\begin{figure*}[t]
\centering
\includegraphics[width=\linewidth]{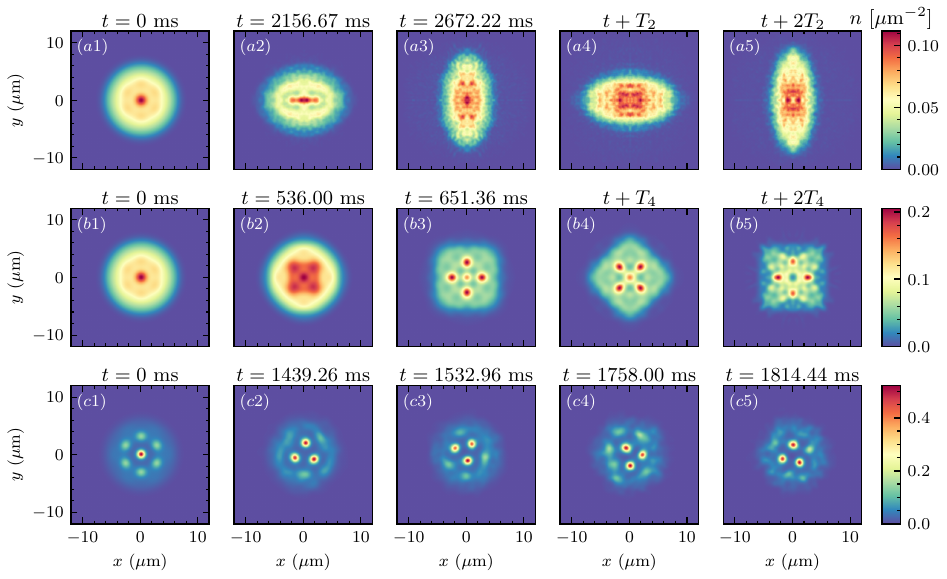}
\caption{Integrated density profiles at different time-instants (see legends) of a driven dipolar gas initialized at the superfluid-to-supersolid phase transition with scattering length $a=95~a_0$. 
The system is under periodic trap modulation governed by Eq.~(\ref{Eq:Driving_trap}) with (a1)-(a5) $\omega_d=2\pi \times 111~\rm{Hz}$ exciting the $\ell=2$ surface mode and (b1)-(b5) $\omega_d=2\pi \times 129~\rm{Hz}$ triggering the $\ell=4$ mode. 
These resonant frequencies are lower compared to the ones extracted in the superfluid regime. 
Both surface and bulk patterns build-up on top of the density of the gas. (c1)-(c5) Density evolution of an initial supersolid state at $a=93~a_0$ subjected to confinement driving with frequency $\omega_d= 132~\rm{Hz}$ . 
The appearance of a bulk triangular pattern is observed which turns out to be robust across a large interval of driving frequencies. 
In all cases the driving amplitude is $\mathcal{A}=0.2$.} 
\label{transitions_patterns}
\end{figure*}

So far, we have investigated the emergence of pattern formation in dipolar superfluids, and explored the influence of dipolar interactions on the resonant frequencies.
In the superfluid regime, however, it is known that short-range interactions dominate over the dipolar ones, as it can be deduced, for instance, by computing the individual energy contributions in the ground state phases~\cite{Halder_control_2022,Mistakidis_tunneling_SS}. 
To further study the impact of dipolar interactions on the resultant  pattern formation induced by periodic driving, we next consider different phases of dipolar gases. Specifically, we first focus on the superfluid-to-supersolid crossover at $a=95~a_0$ [Fig.~\ref{fig:Ground_states}(c)], where dipolar and contact interaction strengths start to become comparable. 
As argued in Sec.~\ref{sec:Protocol}, in this crossover region the system is perturbed by means of confinement parametric driving, which circumvents the dynamical crossing of the underlying phase boundaries and results in milder excitation of the initial state when compared to interaction strength driving. Moreover, the amplitude of the employed confinement driving is slightly reduced to $\mathcal{A}=0.2$ (as compared to scattering length driving) such that the ensuing perturbation is sufficiently weak and avoids strong excitations caused by large energy transfer~\cite{bougas2026generation}.

Surface pattern formation still prevails in the crossover region, as demonstrated e.g. in Fig.~\ref{transitions_patterns}(a1)-(a5) and (b1)-(b5) for the $\ell=2$ and $\ell=4$ surface modes respectively.
All the previously identified surface modes can nevertheless be excited (not shown) at the appropriate driving frequency window [see also Fig.~\ref{fig:fd_vs_scattering_and_sigma}(a)], except from the $\ell=3$ mode, the absence of which requires further future investigation. The different patterns arise at resonant frequencies lower compared to the respective ones in dipolar superfluids, highlighting the importance of the enhanced dipolar interactions, see also Fig.~\ref{fig:fd_vs_scattering_and_sigma}(a).
In line with the dependence of $\omega_d/(2\pi)$ on $a$ [Fig.~\ref{fig:fd_vs_scattering_and_sigma}(a)], the square pattern ($\ell=4$) is affected the most as compared to the elliptical mode; the associated resonant frequency is significantly reduced, c.f. $\omega_d/(2\pi) = 129~\rm{Hz}$ at $a=95~a_0$ and $\omega_d/(2\pi) = 147~\rm{Hz}$ at $a=150~a_0$.
In contrast, the $\ell=2$ surface mode occurs at roughly the same frequency as for $a=150~a_0$.
The stark difference with the pattern excitation in the superfluid regime is the appearance of bulk patterns at the crossover region, in the form of pronounced density peaks [see e.g. Fig.~\ref{transitions_patterns}(a5), (b5)], which assemble in polygonal structures especially for $\ell>2$. These bulk excitations follow the sub-harmonic response of the surface modes, see Fig.~\ref{transitions_patterns}(b3)-(b5).  
The appearance of such bulk  structures is rooted in the onset of the roton instability that manifests in dipolar gases with strong dipolar interactions~\cite{Schmidt_roton_2021,Petter_probing_2019,chomaz2018observation}.

Another important aspect of pattern nucleation in the crossover region is the faster onset of all observed surface modes\footnote{As in the case of dipolar superfluids the presence of initial perturbations, e.g. emulated by a uniform noise distribution, can further accelerate pattern formation in dipolar supersolids.}. 
In particular, the elliptical mode occurs roughly $\sim 1000~\rm{ms}$ faster compared to the respective mode in the superfluid regime [c.f. Fig.~\ref{transitions_patterns}(a2) and Fig.~\ref{fig:spectral_strength_modes} for $\ell=2$].
The acceleration of pattern formation may be related to the fact that small density deformations are already present initially [Fig.~\ref{transitions_patterns}(a1)], due to the pronounced dipolar interactions.
As we already know from the Mathieu framework [see also Eq.~(\ref{Eq:mode_expansion})], such modulations act as a catalyst for the subsequent generation of patterns~\cite{Pikovsky_synchronization_2001}. 
A similar behavior was also observed with the onset of wave turbulence in dipolar superfluids and supersolids~\cite{bougas2026generation}.
The latter expedite the development of turbulent cascades due to their pronounced density structures, and enhanced amplitude of larger momentum modes.

When the scattering length is tuned to even lower values we approach the supersolid regime, where the dipolar interactions start to dominate over the short-range ones. 
In this interaction region, symmetric crystal configurations arise on the ground state, stabilized by the LHY term, as exemplified in Fig.~\ref{transitions_patterns}(c1) pertaining to $a=93~a_0$.
As in the crossover regime, the supersolid is driven out-of-equilibrium by periodically modulating the frequency of the radial trap.
During the initial stages of the evolution, the six outer crystals precess and subsequently melt due to their mutual collisions as well as interactions with their superfluid background~\cite{Mistakidis_tunneling_SS}, creating an excited superfluid halo surrounding the central droplet peak [not shown for brevity].
At longer evolution times three droplet crystals emerge at the center [Fig.~\ref{transitions_patterns}(c2)], which subsequently interact with their  encompassing superfluid background. 
These interaction events  lead to a rearrangement of the three droplets [Fig.~\ref{transitions_patterns}(c3)-(c5)], whose orientation changes over time in a non-periodic fashion. Such configurations are found to be robust up to $t\sim 3000 ~ \rm{ms}$, and persist over a broad range of driving frequencies, $100~\mathrm{Hz} \lesssim \omega_d/2\pi \lesssim 240~\mathrm{Hz}$ .
However, angular surface patterns are absent during the entire dynamics, a behavior that is attributed to the suppressed superfluid background in dipolar supersolids.

The superfluid background plays an important role for the bulk pattern formation in supersolids appearing for $92~a_0 \gtrsim a \gtrsim 87~a_0$.
When the scattering length is lowered even further, at $a=92~a_0$, the superfluid substrate becomes even more suppressed, see also Fig.~\ref{fig:Ground_states}(a), (b). 
As a consequence, following confinement driving, the hexagonal crystal configuration remains almost unperturbed up to $t \sim 2000~\rm{ms}$, displaying small displacements from their equilibrium positions (not shown). 
The superfluid background becomes eventually excited, leading to the destabilization of the original crystal geometry.
In the long time dynamics a four-droplet structure appears at the trap center, encircled by an excited superfluid halo.
The orientation of the four crystals remains almost fixed during time (in contrast to the above-described triangular pattern at $a = 95~a_0$), and the droplets exhibit a small-amplitude sloshing motion. 
Turning to the droplet phase occurring at $a\leq86~a_0$, we observe a drastically different response (not shown for brevity). 
Again, the formation of surface modes is completely inhibited since the superfluid background is entirely absent. 
Rather, the ensuing crystals undergo small-amplitude displacements in the course of the evolution, reflecting the droplet rigidity~\cite{Bougas_signatures_2026,Mukherjee_classical_2023}.

\section{Summary and Conclusions}
\label{summary and conclusion}

The emergence and dynamics of Faraday surface patterns has been investigated in harmonically trapped 3D dipolar quantum gases governed by the eGPE which includes the beyond mean-field LHY energy correction. 
The magnetic atoms are initialized either in the superfluid or the supersolid phase, and are consecutively subject to periodic driving of the scattering length and the trap frequency respectively. 
In the superfluid case, both protocols excite the same patterns at the same resonant frequencies, while it turns out that modulating the confinement is more favorable for inducing patterns in the supersolid regime.

Focusing on the superfluid state, we demonstrate the resonant excitation of $l$-fold polygonal surface modes manifested by characteristic angular density deformations, and displaying sub-harmonic response.  
To understand the origin of the ensuing instability we have derived the corresponding Mathieu equation based on linear stability analysis. Moreover, Floquet theory has been employed in order to determine the respective stability diagram as a function of the driving frequency and amplitude. 
It is shown that the eGPE predictions are in excellent quantitative agreement with the analytical results for strong axial confinement, across all surface modes considered. 
For weaker axial confinement, the agreement persists for the lower order modes, while discrepancies start to become noticeable for higher modes. 
We attribute these deviations, at least in part, to the coupling between the surface-mode dynamics and the axial degrees-of-freedom. 
This effect becomes more relevant as the surface-mode frequency approaches the axial trap frequency, where the single-mode approximation with frozen axial dynamics (underlying our analytical treatment) is invalid.

Moreover, the Mathieu framework allows us to assess
the influence of dipolar interactions on 
the parametric resonance windows. The latter are shifted to lower frequencies compared with those occurring solely from contact interactions. 
Such an effect is highlighted by exploring the dependence of the resonant frequencies on the trap aspect ratio.
In the strong confinement limit, where the axial motion becomes frozen, the effect of the dipolar interaction is suppressed along the radial plane, and the interaction effectively reduces to a renormalized contact coupling~\cite{he2025dipolar}. 
Consequently, the dipolar contribution to the surface restoring force vanishes, and the resonance frequencies systematically increase, saturating to the values pertaining to non-dipolar superfluids. 
This crossover from a dipolar superfluid to an effectively contact interaction regime is further confirmed by the excellent agreement between the analytical predictions and the eGPE simulations. 
We further explore how the resonance frequencies depend on the contact interaction strength. 
Increasing the scattering length shifts the resonance frequencies toward higher values which, is traced back to the fact that the enhanced repulsive contact interactions stiffen the surface of the condensate and hence increase the restoring force of the surface modes.

Overall, this analytical framework is valid within the superfluid phase, but it breaks down near the supersolid transition. 
In the crossover region between these two phases, the response is enriched due to the coexistence of surface and bulk excitations. 
Importantly, the ground states at the crossover are characterized by small density fluctuations due to roton softening, leading to an acceleration of pattern formation as compared to superfluids.
Turning to the supersolid state, the observed dynamics features metastable droplet lattice density structures and surface deformations are suppressed. 
Hence, we can infer that the response of the dipolar gas to external periodic driving transitions from being surface mode dominated in the superfluid regime, to a hybrid of surface and bulk modes at the superfluid-to-supersolid transition and bulk mode prevailing at the supersolid state.

Our results highlight
the nucleation of parametrically excited surface modes in long-range interacting quantum fluids paving the way for a plethora of future investigations aiming to study the manifestation of  instabilities in these systems. 
A straightforward one is to analyze the emergent  pattern formation and in particular the impact of dipolar interaction at the intersection of the Floquet tongues where nonlinear mode mixing is expected to occur~\cite{kwon2021spontaneous}. 
Similarly, triggering bulk excitations at larger driving frequencies in order to create synthetic supersolids is an especially exciting frontier~\cite{brakensiek2026generation}. 
Additionally, it would be valuable to develop linear stability analysis corroborated by variational calculations in order to understand the role of roton excitations and their coupling to surface Faraday modes at the superfluid-to-supersolid crossover~\cite{santos2003roton,chomaz2018observation}. Along these lines, using parametric driving to probe the roton spectrum and in particular its softening and momentum modes is desirable~\cite{Nath2010,Lakomy2012}. 
Another promising direction concerns the study of nonequilibrium phase transitions and defect formation induced by periodic driving in dipolar supersolids~\cite{tanzi2019observation,bottcher2019transient,bougas2026generation}. 
Finally, it would also be interesting to explore instability induced pattern formation in two-component dipolar gases~\cite{Trautmann_dipolar_2018,Lecomte_production_2025,Duerbeck_dipolar_2026} where mixed superfluid-supersolid and supersolid states of matter coexist~\cite{Bland_alternating_2022} and also interfacial modes~\cite{maity2020parametrically} can be potentially excited.

\section*{Acknowledgments} 
S.I.M acknowledges support by the Army Research Office under Award number: W911NF-26-1-A043.

%\clearpage
\appendix

\renewcommand{\thesection}{\Alph{section}} 
\counterwithin{figure}{section}
\renewcommand{\theequation}{\thesection.\arabic{equation}} 
\setcounter{equation}{0}  
\section{Equivalence between scattering length and radial trap modulation}
\label{app:scattering}

In binary immiscible superfluids it was argued that Faraday resonances can be equivalently excited using either interaction strength or trap frequency modulation~\cite{maity2020parametrically}. Here, we numerically showcase that the same phenomenology holds in the case of the considered dipolar superfluid.

\begin{figure}[!htp]
    \centering
    \includegraphics[width=\linewidth]{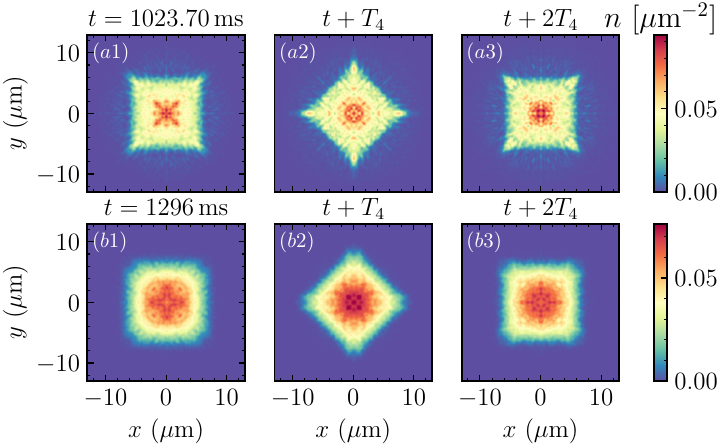}
  \caption{
  Equivalence between the interaction (a1)-(a3) and confinement (b1)-(b3) modulation in the case of a dipolar superfluid initialized at $a_i=150~a_0$. In both cases the dipolar quantum gas is driven at $\omega_d/(2\pi)= 147~\rm{Hz}$, leading to the appearance of the $\ell=4$ surface mode.
}
    \label{fig:four_lobe_trap}
\end{figure}

To corroborate this equivalence, 
we assume the excitation of the $\ell=4$ surface mode of a superfluid at $a_i=150~a_0$, by employing either the interaction [Fig.~\ref{fig:four_lobe_trap}(a1)-(a3)] or the confinement modulation [Fig.~\ref{fig:four_lobe_trap}(b1)-(b3)] protocol.
In both cases the same four-lobed pattern emerges for a frequency $\omega_d/(2\pi) = 147~\rm{Hz}$, implying that the subsequent precession dynamics is identical in both scenarios.
A difference that occurs between the two protocols is in the pattern formation time; modulating the scattering length leads to a slightly faster onset of pattern formation as compared to the periodic driving of the trap [c.f. Fig.~\ref{fig:four_lobe_trap}(a1) and (b1)].
This suggests that the effective modulation strength differs between the two protocols. A more rigorous proof of the equivalence between the two protocols is a fruitful direction for future study.
Moreover, we have numerically confirmed that all other surface modes can be equally excited in dipolar superfluids with either the radial trap or the scattering length modulation.
The only exception is for $\ell=3$, which appears only upon driving the short-range interactions. The absence of this pattern in the case of radial trap modulation requires further investigation.

%%%%%%%%% %============================================================
% APPENDIX
%============================================================
\section{Derivation of the Mathieu equation}
\label{app:mathieu}

Here, we 
provide the detailed derivation of the Mathieu equation governing the dynamics of small surface deformations of the dipolar condensate. 
The starting point is the linearized density wave equation obtained via the Madelung transformation (Sec.~\ref{sec:Mathieu}),
\begin{equation}
m
\frac{\partial^2 \delta\rho}{\partial t^2}
=
\boldsymbol{\nabla}\!\cdot\!
\left[
\rho_0
\boldsymbol{\nabla}
\Big(
g(t)\,\delta\rho
+
\Phi_{\rm{dd}}[\delta\rho]
\Big)
\right],
\label{eq:linear_density_wave_app}
\end{equation}
where $g(t)=g_0\left[1+\mathcal{A}\cos(\omega_d t)\right]$ and $g_0=\frac{4\pi\hbar^2  a_i}{m}$.
To account for the appearance of $\ell$-fold patterns, we assume deformations of the form,
\begin{equation}
\delta\rho(r,\phi,t)
=
\zeta_\ell(t)\,r^\ell\cos(\ell\phi).
\label{Eq:mode_expansion_app}
\end{equation}
where 
$(r,\phi)$ are the polar coordinates.
This ansatz is harmonic i.e. $\nabla^2 \delta \rho = 0$, and thus the first term on the right hand side of Eq.~(\ref{eq:linear_density_wave_app}) simplifies to $\boldsymbol{\nabla}\!\cdot\!(\rho_0 \boldsymbol{\nabla} \delta\rho)=\boldsymbol{\nabla}\rho_0 \cdot \boldsymbol{\nabla} \delta\rho$.
To evaluate the latter term we employ the equation for the superfluid velocity at equilibrium [Eq.~\eqref{eq:velocity derivative}], i.e. $g_0\boldsymbol{\nabla} \rho_0 = - \boldsymbol{\nabla} V  - \boldsymbol{\nabla} \Phi_{\rm{dd}}[\rho_0] $.

To handle these terms,
the equilibrium configuration, $\rho_0$, is approximated by a Thomas-Fermi density profile~\cite{odell2004exact,Eberlein2005}, $\rho_0=\rho_{0c} \left( 1-r^2/R_r^2 - z^2/R_z^2  \right)$, where $R_r$ ($R_z$) is the Thomas-Fermi radius along the planar (axial) direction, and $\rho_{0c} = \frac{15N}{8\pi R_r^2 R_z}$.
The radii are determined by the following expressions,
\begin{subequations}
\begin{eqnarray}
R_r & =&
\left[
\frac{15\,g_0\,N\,\kappa}{4\pi m \omega_r^2}
\left(
1+\varepsilon_{\mathrm{dd}}
\left(
\frac{3\kappa^2 f(\kappa)}{2(1-\kappa^2)}-1
\right)
\right)
\right]^{1/5},
\label{Rx} \nonumber \\ \\
R_z & = & R_r/\kappa,
\label{R_z} \\
f(\kappa) & = &\frac{2+\kappa^{2}\left[4-\frac{6\tan^{-1}(\sqrt{\kappa^2-1})}{\sqrt{\kappa^2-1}}\right]}{2\left(1-\kappa^{2}\right)}.
\end{eqnarray}
\end{subequations}
Moreover, the condensate aspect ratio, $\kappa$, is obtained by solving the following transcendental equation,
\begin{equation}
3\kappa^2\varepsilon_{\mathrm{dd}}
\left[
\left(\frac{\gamma^2}{2}+1\right)
\frac{f(\kappa)}{1-\kappa^2}-1
\right]
+
(\varepsilon_{\mathrm{dd}}-1)(\kappa^2-\gamma^2)=0,
\label{f(k)}
\end{equation}
with $\gamma=\omega_z/\omega_r$ the trap aspect ratio.
Importantly, the Thomas-Fermi approximation leads to an analytic expression for the dipolar potential in the ground state~\cite{odell2004exact,Eberlein2005},
\begin{eqnarray}
\Phi_{\mathrm{dd}}[\rho_0]&=&
\frac{15g_0N \varepsilon_{\mathrm{dd}}}{8\pi R_r^2 R_z}
\Bigg[
\frac{r^2}{R_r^2}
-\frac{2z^2}{R_z^2} \nonumber \\
& &  -f(\kappa)
\left(
1-\frac{3}{2}
\frac{r^2-2z^2}{R_r^2-R_z^2}
\right)
\Bigg],
\label{Eq:TF_approximation}
\end{eqnarray}

Having at hand this expression, the first term on the right hand side of Eq.~\eqref{eq:linear_density_wave_app} can be written as
\begin{subequations}
\begin{gather}
    \boldsymbol{\nabla} \rho_0 \cdot \boldsymbol{\nabla} \delta \rho = \{ - \boldsymbol{\nabla} V - \boldsymbol{\nabla} \Phi_{\rm{dd}}[\rho_0] \} \cdot \boldsymbol{\nabla}\delta \rho / g_0 \nonumber \\
    = - (m \omega_r^2 + 2 A_{\rho}) \ell \delta \rho / g_0, \label{Eq:Contributions_right_hand_side} \\
    A_\rho = g_0 \rho_{0c} \epsilon_{\rm{dd}} \left(\frac{1}{R_r^2} + \frac{3f(\kappa)}{2(R_r^2-R_z^2)}\right). \label{eq:A_rho_explicit}
\end{gather}
\end{subequations}
Therefore, Eq.~\eqref{Eq:Contributions_right_hand_side} is decomposed into two contributions. 
The first term stems from the trapping potential and is identical to the term appearing in non-dipolar superfluids~\cite{kwon2021spontaneous}.
The second one, $A_{\rho}$ stems from the dipolar potential at equilibrium.

The second contribution to Eq.~(\ref{eq:linear_density_wave_app}) is the dipolar potential generated by the perturbation itself,
\begin{equation}
\Phi_{\rm{dd}}[\delta\rho]
=
\int d\mathbf{r}' ~ U_{\rm{dd}}(\mathbf{r}-\mathbf{r}')\,\delta\rho(\mathbf{r}').
\end{equation}
For polynomial perturbations of degree $\ell$ on an ellipsoidal domain, this integral is guaranteed to remain proportional to $\delta \rho$. This is a consequence of the Ferrers/Dyson degree theorem together with conservation of angular momentum about the polarization axis~\cite{dyson1891potentials}. Hence,
\begin{equation}
\Phi_{\rm{dd}}[\delta\rho]
=
B_\ell\,\delta\rho,
\end{equation}
where the coefficient $B_\ell$ follows from the alternative form of the dipolar potential~\cite{van2010collective},
\begin{subequations}
\begin{eqnarray}
\Phi_{\rm{dd}}[\rho]
& = &
- 3 g_0 \epsilon_{\rm{dd}}
\left(
\frac{\partial^2 v[\rho]}{\partial z^2}
+
\frac{\rho}{3}
\right),
\label{eq:Phi_dd} \\
v[\rho]
& = &
\frac{1}{4\pi}
\int d\mathbf{r}' ~
\frac{\rho(\mathbf{r}')}{|\mathbf{r}-\mathbf{r}'|}.
\label{eq:Phi_aux}
\end{eqnarray}
\end{subequations}
In particular, since $\frac{\partial^2}{\partial z^2}v\bigl[r^\ell\cos(\ell\phi)\bigr]
=
C_\ell\,r^\ell\cos(\ell\phi)$,
it holds that
\begin{equation}
B_\ell
=
-3 g_0 \epsilon_{\rm{dd}}
\left(
C_\ell
+
\frac{1}{3}
\right).
\end{equation}
The constants $C_\ell$ are evaluated using the closed-form polynomial-potential expressions, and the associated shape integrals of Ref.~\cite{van2010collective}, see also below for further details.

Because $\Phi_{\rm{dd}}[\delta\rho]$ is proportional to $\delta\rho$ with exactly the same spatial structure as the contact term, the second term on the right hand side of Eq.~\eqref{eq:linear_density_wave_app} becomes
\begin{equation}
\boldsymbol{\nabla}\!\cdot\!\bigl(\rho_0 \boldsymbol{\nabla} \Phi_{\rm{dd}}[\delta\rho]\bigr)
=
B_\ell \,\boldsymbol{\nabla}\!\cdot\!(\rho_0 \boldsymbol{\nabla} \delta\rho)
=
- \frac{2\rho_{0c}}{R_r^2}\, B_\ell \,\ell \,\delta\rho.
\label{Eq:Second_term}
\end{equation}
Defining, in direct analogy with $A_\rho$,
\begin{equation}
A_\ell
\equiv
\frac{\rho_{0c}}{R_r^2}\, B_\ell,
\end{equation}
Eq.~\eqref{Eq:Second_term} reduces to $-2 A_\ell\, \ell \,\delta\rho$.

Collecting both contributions into Eq.~(\ref{eq:linear_density_wave_app}), and using the ansatz for $\delta\rho$ [Eq.~\eqref{Eq:mode_expansion}],
we arrive at the following Mathieu equation
\begin{subequations}
\begin{gather}
\frac{d^2\zeta_\ell}{dt^2}
+
\omega_\ell^2
\left[
1
+
\frac{\ell\left(\omega_r^2+\dfrac{2A_\rho}{m}\right)}
{\omega_\ell^2}
\,\mathcal{A}\cos(\omega_d t)
\right]
\zeta_\ell
=0,
\label{eq:mathieu_final_app} \\
\omega_{\ell}^2
=
\ell\,\omega_r^2
+
\ell\,\frac{2A_\rho}{m}
+
\ell\,\frac{2A_\ell}{m}.
\label{eq:omega_app}
\end{gather}
\end{subequations}

In the following, we illustrate the calculation of $A_\ell$ for the case $\ell=2$. The density deformations are proportional to $r^2\cos(2\phi)=x^2-y^2$. Subsequently, we consider separately the terms $\partial_z^2v[\rho_1]$ and $\partial_z^2v[\rho_2]$, where $\rho_1 = x^2$, $\rho_2=y^2$. 
Following the same procedure as in Ref.~\cite{van2010collective}, these two terms can be expressed in terms of polynomials,
\begin{subequations}
\begin{gather}
\partial_z^2v[\rho_1] = 0.2742\,z^2+0.0378\,y^2-0.3120\,x^2, \\
\partial_z^2v[\rho_2] = 0.2742\,z^2+0.0378\,x^2-0.3120\,y^2.
\end{gather}
\end{subequations}
Subtracting these terms yields
\begin{equation}
\partial_z^2\phi[\rho_1-\rho_2] = -0.3497\,(x^2-y^2),
\end{equation}
which leads to $C_2=-0.3497$. For $\omega_z=2\pi\times133$~Hz, where $\kappa=2.2657$, $R_x=8.874$~$\mu$m, $R_z=3.916$~$\mu$m, $n_{0c}=1.548\times10^{20}$~m$^{-3}$, this corresponds to $A_2=+3.444\times10^{-22}~\text{J}\cdot\text{m}^{-3}$.
Combining this with the suitable background coefficient $A_\rho=-1.930\times10^{-21}~\text{J}\cdot\text{m}^{-3}$ [Eq.~\eqref{eq:A_rho_explicit}], Eq.~\eqref{eq:omega_app} yields $\omega_2/2\pi=55.75$~Hz, in agreement with the value quoted in Sec.~\ref{sec:Mathieu}.

%%%%%%%%%%%Appendix-3%%%%%%%
\section{Surface modes of dipolar superfluids in the strong transverse confinement limit}
\label{app:strong}
\setcounter{equation}{0}

In this appendix, we derive the asymptotic form of the surface-mode frequencies in the limit of strong transverse confinement. In this regime, the dipole--dipole interaction becomes effectively local, allowing the general Mathieu dispersion relation derived in Appendix~\ref{app:mathieu} to reduce to the well-known hydrodynamic result for condensates with short-range interactions. 

When the axial confinement energy greatly exceeds the chemical potential, $\hbar\omega_z \gg \mu$, the dipolar condensate enters the two-dimensional regime, where the axial motion is effectively frozen. In this limit, the condensate is locally uniform in the transverse plane, and the dipole--dipole interaction reduces to an effective contact interaction~\cite{Parker_Thomas_2008}, $\Phi_{\mathrm{dd}}[\rho] = 2g_0\varepsilon_{\mathrm{dd}} \rho$. This is equivalent to replacing the $s$-wave scattering length by the effective value $a_{\mathrm{eff}} = a\,(1 + 2\varepsilon_{\mathrm{dd}})$. To leading order, the long-range dipolar interaction is therefore absorbed into a renormalized local coupling.
Consequently, we recover the well-known hydrodynamic result $\omega_{\ell}=\sqrt{\ell}\,\omega_r$ for condensates with purely contact interactions~\cite{kwon2021spontaneous}.
As an example, for $\omega_z/(2\pi) = 4100~\rm{Hz}$, the corresponding natural frequencies in the limit $\mathcal{A} \to 0$ are $\omega_{\ell}/(2\pi) \approx 60.63,\,74.07,\,85.35,\,95.24,$ and $104.14~\mathrm{Hz}$ for the $\ell = 2,3,4,5,6$, respectively. These analytical predictions are in excellent agreement with the corresponding eGPE simulations in the strong-transverse-confinement regime [see Fig.~\ref{fig:fd_vs_scattering_and_sigma}(a)], further confirming the reduction of the dipolar system to the effective contact-interaction limit.

Alternatively, the dispersion relation in the strong-confinement limit can be extracted from Eq.~\eqref{eq:omega_app}. In particular, it can be shown that the combined dipolar terms 
\begin{equation}
\frac{2A_\rho}{m} + \frac{2A_\ell}{m}
\approx 0,
\qquad (\hbar\omega_z \gg \mu),
\label{eq:cancellation_app}
\end{equation}
independently of the azimuthal quantum number $\ell$ and the dipolar interaction strength $\varepsilon_{\mathrm{dd}}$. As such, it holds that $\omega_{\ell}^{2} \approx \ell\,\omega_r^{2}$.

\bibliography{references}
%\bibliography{reference}

\end{document}